\documentclass[aps,prl,twocolumn,showpacs,superscriptaddress]{revtex4}

\usepackage{amsmath}
\usepackage{amsfonts}
\usepackage{amssymb}

\newcommand{\beqa}{\begin{eqnarray}}
\newcommand{\eeqa}{\end{eqnarray}}
\newcommand{\beq}{\begin{equation}}
\newcommand{\eeq}{\end{equation}}

\begin{document}

\title{Reducing Polarization Mode Dispersion With Controlled Polarization Rotations}

\author {Serge Massar}
\email{smassar@ulb.ac.be}
\affiliation{Laboratoire d'Information Quantique and Centre for Quantum Information and Communication, {C.P.} 165/59, Universit\'{e} Libre de Bruxelles, Avenue F. D. Roosevelt 50, 1050 Bruxelles, Belgium}
\author{S. Popescu}
\affiliation{H. H. Wills Physics Laboratory, University of Bristol, Tyndall 
      Avenue, 
      Bristol BS8 1TL, U.K.}
\affiliation{Hewlett-Packard Laboratories, Stoke Gifford, Bristol BS12
  6QZ, U.K.}

\begin{abstract}

One of the fundamental limitations to high bit rate, long distance, telecommunication in optical fibers is Polarization Mode Dispersion (PMD). Here we introduce a conceptually new method to reduce PMD in optical fibers by carrying out controlled rotations of polarization at predetermined locations along the fiber. The distance between these controlled polarization rotations must be less than both the beat length and the mode coupling length of the fiber. This method can also be combined with the method in which the fiber is spun while it drawn. The incidence of imperfections 
on the efficiency of the method is analysed.

\pacs{42.81.Gs 42.81.Bm}

\end{abstract}

\maketitle


{\bf Introduction.}
Polarization Mode Dispersion (PMD) arises when two orthogonal states of polarization propagate at different velocities  in optical fibers. PMD is one of the fundamental limitations on high speed, high bit rate communication in fiber systems because it distorts the shape of light pulses, and in particular induces pulse spreading. Hence much effort has gone into reducing PMD in optical fibers, see \cite{LPMD,Book} for reviews.

PMD arises because of uncontrolled stresses or anisotropies induced in the fiber during the manufacturing process and during deployement. These cause unwanted birefringence, and hence PMD. This residual birefringence changes randomly along the fiber, resulting in random mode coupling as the light propagates along the fiber. Because of this statistical process, the effects of PMD -such as pulse spreading- increase as the square root of the propagation distance\cite{Poole88,Poole89,PHM,gisin1,gisin2}. 
This statistical process makes it extremely difficult to correct the effects of PMD after the light has propagated through a long length of fiber. For this reason one rather tries to reduce the PMD of the fiber itself.

Two main methods have been devised to reduce PMD in optical fibers. The first is to minimize asymmetries in the index profile and stress profile of the fiber. To this end the manufacturing process has been steadily improved. 
The second method is to spin the fiber during the manufacturing process as described in \cite{Barlow,spinningpatent}. Spinning does not reduce the anisotropies in the fiber. Rather it modifies the orientation of the anisotropies along the fiber in such a way that after one spin period the effective birefringence and PMD is reduced. The effects of spinning on PMD were analyzed in e.g. \cite{li1,sss,li2,li3}. 
Residual birefringence can be characterized by the fiber's beat length, which in present day fibers is of order 10m.

In the present work we introduce a conceptually different method for reducing PMD in optical fibers. Our basic idea is to introduce controlled polarization rotations at predetermined locations along the fiber in such a way that the effects of PMD are reduced. This is paradoxical since the aim is to reduce birefringence and PMD, and we claim to achieve this by introducing additional polarization rotations. Our main idea is that after several such controlled polarisation rotations, the state of polarisation of the light has been oriented along many different directions on the Bloch sphere in such a way that the effects of the fiber birefringence averages out. We will show that this can be a highly effective method. It can also be combined with the method of spinning the fiber.

We will not be concerned here with how the controlled polarization rotations are implemented in practice. We note that there are many ways of inducing controlled birefringence in fibers, and hence controlled polarization rotations. Typically these imply inducing  stress in the fiber, for instance through compression, twisting, etc.  We leave for future work to determine which implementation of controlled polarization rotations is most suitable for reducing PMD.

The method presented here for reducing PMD in optical fibers is inspired by the 
pulse sequences (e.g. spin echo)
that have been developed in Nuclear Magnetic Resonance (NMR) to 
reduce the effects of imperfections, see for instance \cite{nmr}. It is also related to the "bang-bang" technique introduced by Viola and Lloyd to reduce decoherence in the context of quantum information\cite{VL,VKL}. Wu and Lidar \cite{WL} suggested applying bang-bang control to reduce losses in optical fibers through controlled phase shifts, but their proposal is impractical because the control operations need to be implemented over distances comparable to the wavelength of light. On the other hand in the present case the control operations need to be implemented over distances of order 1m, which makes reducing PMD with controlled polarisation rotations a practical proposal. An important conceptual difference which we stress is that in the bang-bang technique one wishes to average to zero the coupling between the system and the environment, whereas in the present case it is the coupling between the polarisation and frequency of light itself, ie. two degrees of freedom of the same system, which one wishes to average to zero.


{\bf Polarized light propagating in optical fibers.}
Consider a light pulse propagating along a birefringent optical fiber. The pulse is centered
 on frequency $\Omega$ and has wave number $K$. The distance along the fiber is
 denoted $l$. Its amplitude can be written as
\begin{equation} 
A(l,t)e^{-i \Omega t + i Kl}
\end{equation}
 where the slowly varying envelope of the pulse $A(l,t)=\left(\begin{array}{c}{A_1(l,t)}\\{A_2(l,t)}\end{array}\right)$ is a two component vector, called the Jones vector, which describes the polarisation state of the light. We introduce the variable $t' = t -
l/v_g$ where $v_g$ is the group velocity of the pulse, whereupon $A$ obeys the evolution equation
\begin{equation} i
\partial_l A =  B_0(l) A + i  B_1(l) \partial_{t'} A 
\label{At}
\eeq where $B_0$ and $B_1$ are traceless, hermitian matrices describing the birefringence and PMD along the fiber. (The parts of $B_0$ and $B_1$ proportional to the
identity are incorporated into the wave vector $K$ and the
average group velocity $v_g$. For simplicity we neglect loss, dispersion, and non linearities.)
The term proportional to $B_0$ describes the phase birefringence.
Its magnitude is generally measured by the beat length $L_B= \pi / \vert B_0\vert$ which is the length after which the polarization is brought back to its original state. If only $B_0$ was present in the fiber, the state of polarization would evolve in an unknown way, but there would be no pulse spreading. The PMD is encoded in the group birefringence $B_1$ which describes the difference in group velocities of the two orthogonal states of polarization. This term will give rise to spreading of pulses. The birefringence changes along the fiber, hence both $B_0(l)$ and $B_1(l)$ depend on the position $l$ along the fiber. In present day low birefringence fibers both $L_B$ and the distance over which $B_0$ and $B_1$ change are comparable and of order $10m$.
 
We can rewrite these matrices as
\begin{equation}
B_0 (l) = \vec B_0(l) \cdot \vec \sigma\quad 
\mbox{and}\quad 
B_1(l) = \vec B_1(l) \cdot \vec\sigma
\label{BBBB}
\end{equation}
where 
$\sigma_x=\left(\begin{array}{cc}0&1\\1&0\end{array}\right)$, 
$\sigma_y = \left(\begin{array}{cc}0&-i\\i&0\end{array}\right)$, 
$\sigma_z = \left(\begin{array}{cc}1&0\\0&-1\end{array}\right)$ are the Pauli matrices,
and $x,y,z$ denote orthogonal directions on the Poincaré sphere.
Denoting by $A (l,\omega)$ the Fourier transform of $A(l,t')$ with respect to $t'$, eq. ( \ref{At}) becomes
\begin{eqnarray}
&i \partial_l A (l,\omega) =
\vec B(l,\omega) \cdot \vec \sigma A(l,\omega)\ ,&
 \label{B}
\\
&\mbox{ where }\quad
\vec B (l,\omega) = \vec B_0(l)  + \omega \vec B_1(l) & \label{BB}
\end{eqnarray}
is the sum of the birefringence and PMD in the fiber.
The solution of eq. (\ref{B}) can be expressed as
\begin{equation}
A(l,\omega) = U(l,\omega) A(0,\omega)\label{AU}
\end{equation}
with $U(l)$ a $2\times 2$ unitary matrix, called the Jones matrix which describes 
the evolution  of the state of polarization after propagating a distance $l$


{\bf Basic Principle of the method.}
For simplicity we will first suppose in the following paragraphs that $\vec B(\omega)$ is
independent of $l$, whereupon the 
Jones matrix is
\beq U(l)=e^{-i\vec B (\omega)\cdot \vec\sigma l}.
\label{timeevol}\eeq
Let us now show that, even though we do not know $\vec B (l,\omega)$, it is possible to
compensate for the evolution so
that after compensation the polarization comes back to its initial state. The
following procedure consisting of four basic steps that are then repeated, realises this:
\begin{enumerate}
\item The light propagates over distance  $l$. The
distance $l$ is taken short enough such that $\vert B \vert
l << 1$, hence the evolution can be well approximated by the
first-order approximation:
\beq U(l)\approx 1-i\vec B \vec\sigma l=1-il(B_x\sigma_x+B_y\sigma_y+B_z\sigma_z)\ .
\nonumber\eeq
\item We interrupt the evolution by flipping the polarization
around the $x$ axis.  
We then let the light evolve over a new distance
$l$ and finally we flip again the polarization around the $x$ axis. The
evolution in step 2 is thus described by $\sigma_x U(l)\sigma_x$. To first order in $\vert B \vert l$, one finds 
\beq \sigma_x
U(l)\sigma_x\approx \sigma_x( 1-i\vec B \vec\sigma
l)\sigma_x=1-il(B_x\sigma_x-B_y\sigma_y-B_z\sigma_z)\nonumber\eeq
which effectively compensates the evolution due to the $B_y$ and
$B_z$ components. 
The evolution due to the $B_x$ component is not
yet compensated; this will be accomplished during the next two
steps.
\item The same as Step 2, but the spin is flipped around the $y$
axis. Step 3 is thus described by $\sigma_y U(l)\sigma_y$. By
expanding $U(l)$ to first order we obtain \beq \sigma_y
U(l)\sigma_y\approx \sigma_y( 1-i\vec B \vec\sigma
l)\sigma_y=1-il(-B_x\sigma_x+B_y\sigma_y-B_z\sigma_z)\nonumber\eeq
\item The same as Step 2, but the spin is flipped
around the $z$ axis. Step 4 is thus described by $\sigma_z
U(l)\sigma_z$. To first order we obtain \beq \sigma_z
U(l)\sigma_z\approx \sigma_z( 1-i\vec B \vec\sigma
l)\sigma_z=1-il(-B_x\sigma_x-B_y\sigma_y+B_z\sigma_z)\nonumber\eeq
\end{enumerate}

Putting all together, the time evolution over the four steps is
\beqa U (l_{seq})&=&\sigma_z U(l)\sigma_z\sigma_y U(l)\sigma_y\sigma_x
U(l)\sigma_xU(l)\label{sssss}
\\ &=& 1 + O(B^2 l_{seq}^2)\label{sss2}
\end{eqnarray} 
where we denote by $l_{seq}=4l$ the length of the sequence.
The evolution is therefore effectively stopped. 
The procedure is then repeated
again and again. Note that because $\sigma_y \sigma_z = i \sigma_x$, eq. (\ref{sssss}) simplifies to
\begin{equation}
U (l_{seq})=\sigma_z U(l)\sigma_x U(l)\sigma_z
U(l)\sigma_xU(l) \ .\label{seqsimple} 
\end{equation}

The method relies on the fact that the interaction, although
unknown to us, does not change along the fiber (at least for the short lengths
$l_{seq}$ we are considering here). The unknown birefringence $\vec B$ that is
responsible for the rotation of the polarization
 in the first place is there all the time, and affects the polarization after
 the $x$, $y$ and $z$ rotations, and brings it back to its initial state at the end of the sequence.

In optical fibers the phase birefringence is of the same order as the group birefringence. This implies that $B_0 / \Omega \simeq B_1$. Since $\omega$, the frequency spread of the pulse, is much smaller than $\Omega$, the carrier frequency, we can expand eq. (\ref{sss2}) to first order in $\omega B_1$ to obtain
$U(l_{seq})= 1 + O(B_0^2 l_{seq}^2) + O( \omega B_1  B_0 l_{seq}^2)$. 
Consider now the evolution over a fixed length $L>l_{seq}$. It is given by
$U(L)= U(l_{seq})^{L/l_{seq}}\simeq 1+ O(B_0^2 l_{seq} L)+O(\omega B_1 B_0 l_{seq} L)$ whereas in the absence of the control sequence it would have been given by $1+ B_0 L + \omega B_1 L$. Hence there is a reduction of PMD by a factor $B_0 l_{seq} = \pi l_{seq}/L_{B}$.

Note that as shown in \cite{VKL} it is possible to devise sequences that cancel not only the terms of order $Bl$, but higher order terms as well. For instance the sequence
\beqa
U_2 (l_{seq})&=&U(l)\sigma_x U(l)\sigma_z
U(l)\sigma_xU(l)U(l)\sigma_x U(l)\sigma_z\times\nonumber\\
& &\times U(l)\sigma_xU(l)
= 1 + O(B^3 l_{seq}^3) \label{SEQ2}
\eeqa 
 cancels the evolution up to order $B^3l^3$. 


{\bf Combining Controlled Polarization Rotations and Spun Fibers.}
It is possible to reduce PMD in optical fibers by combining the methods of controlled polarization rotations and of spinning the fiber while it is drawn. Here we first show how to include both methods in a unified description, and then we discuss how they can be used in combination.

For spun fibers the evolution equation is no longer given by eq. (\ref{B}), but by
\beq
i \partial_l A = U_{spin}^\dagger (l) B(\omega) U_{spin}(l) A
\label{Aspin}
\eeq
where
\beq
U_{spin}(l)=e^{-i \alpha(l) \sigma_y}
\label{Uspin}
\eeq
describes the spin of the fiber, $\alpha(l)$ is the spin function (the angle by which the fiber is spun), and $\sigma_y$ is the Pauli matrix that generate polarization rotations around the circular polarization axis.  To first order in $B$ the evolution is given by
$U(l) = 1 + i \int_0^l dl' U_{spin}^\dagger (l') B(\omega) U_{spin}(l')$, from which one deduces, see \cite{li2}, that in order to reduce PMD the spin function must obey
\beq \int_0^L dl \cos[\int_0^l 2\alpha(l')dl']=0
\mbox{ and }
\int_0^L dl \sin[\int_0^l 2\alpha(l')dl']=0
\nonumber
\eeq  
where $L$ is the spin period (the period of $\alpha(l)$).
Note that spinning the fiber alone can reduce linear birefringence, but not circular birefringence (proportional to $\sigma_y$).

If we incorporate controlled polarization rotations, we get the evolution equation:
\beq
i \partial_l A = \left( U_{spin}^\dagger (l) B(\omega) U_{spin}(l) + B_c(l)\right) A
\label{Aspin2}
\eeq
where $B_c(l)$ (proportional to Dirac delta functions) describes the controlled polarization rotations. It is convenient to rewrite this equation in the frame that rotates with the spin. To this end we define
$\tilde A (l) = U_{spin}(l) A(l)$.
The rotated Jones vector obeys the evolution equation
\beq
i \partial_l \tilde A = \left(  B(\omega)  + \partial_l \alpha(l) \sigma_y+ U_{spin}(l) B_c(l)U_{spin}^\dagger (l)\right) \tilde A\ .
\label{Aspin3}
\eeq
Thus in the rotating frame, spin is formally identical to a continuous controlled polarization rotation given by $ \partial_l \alpha(l) \sigma_y$, whereas the controlled polarization rotations $B_c(l)$ are rotated and take the form $ U_{spin}(l) B_c(l)U_{spin}^\dagger (l)$. Therefore even though they have completely different origins, they can easily be combined. For instance one easily checks that by carrying out a $\sigma_x$ flip of polarization after each spin period one cancels both linear and  circular birefringence.


{\bf Imperfections.}
In the above calculations we have supposed that the 
unknown polarization rotation $\vec B$ does not change along the fiber. The method stays valid if $\vec B(l)$ changes slowly along the fiber. 
To see this suppose that $\vec B(l)$ is a smooth function which can be expanded in  Taylor series
\beq
\vec B (l)= \vec B (0) + l \partial_l \vec B (0) + O (l^2)\ .
\label{series}
\eeq
We define the length over which the birefringence changes by
\beq L_{chge} = \frac{|B(0)|}{|\partial_l B(0)|}\ .
\eeq

Because $B(l)$ is now position dependent, the evolution eq. (\ref{timeevol}) must be replaced by
\beqa
U(l)
&\simeq&
e^{-i \vec B(0) \vec\sigma l} \left (
1 - i \int_0^l d l'\! l'
e^{+i l' \vec B(0) \vec\sigma} \partial_{l'} \vec B (0) \vec \sigma e^{-i l' \vec B(0) \vec\sigma}
\right)
\nonumber\\
&=& e^{-i \vec B(0) \vec\sigma l} \left ( I + O(l^2 \partial_l B )\right)\ .
\label{Uchange}
\eeqa
This expression must be inserted into eq. (\ref{sssss}). The terms proportional to $l^2 \partial_l B (0)$ will not cancel, so after the control sequence, one will be left with
the evolution 
\beqa
U(l_{seq})&=&I + O(  B^2 l_{seq}^2) + O( l_{seq}^2 \partial_l B)
\label{U2}
\nonumber\\
&=&I + O(  \omega^0) +
 O( \omega B_1 B_0 l_{seq}^2) + O( \omega B_1 l_{seq}^2 / L_{chge})
\nonumber
\eeqa
where 
we have kept the terms of first order in $\omega$, and we have supposed that $\partial_l B_1 / B_1 \simeq \partial_l B_0 / B_0 = L_{chge}^{-1}$.
Note that in modern fibers with low birefringence the beat length $L_B$ is comparable to the length $L_{chge}$ over which the birefringence changes. Hence the two terms in 
eq. (\ref{U2}) are of comparable magnitude and there may not be much to gain by going to a sequence such as eq. (\ref{SEQ2}) which cancels the PMD to second order 

The second source of imperfections we consider is the fact that in general the controlled polarisation rotations will not be achromatic, and will therefore themselves introduce PMD. Thus, instead of carrying out the rotation $\sigma_x$, one carries out the rotation $\exp[i \sigma_x \pi (1+ \omega \beta)/2]$ where $\beta \simeq B_1 / B_0 \simeq 1/ \Omega$. In order for the method to be effective we must cancel the PMD due to $\beta$.
To do this we propose using the idea of  
stacking several chromatic waveplates to yield an essentially achromatic waveplate \cite{Achr1}. For instance the following stack of 3 $\pi$ rotations, with $C= 1/2$ and $S=\sqrt{3}/2$, cancels all PMD to order $(\omega\beta)^2$:
\beqa
&e^{i \sigma_x(\pi + \omega \beta)/2}
e^{-i (C \sigma_x - S \sigma_y )(\pi + \omega \beta)/2}
e^{i \sigma_x(\pi + \omega \beta)/2}&\nonumber\\
&=  i (C \sigma_x + S \sigma_y ) + O(\omega \beta)^2\ .&
\eeqa
Thus, by replacing each controlled polarisation rotation by three controlled polarisation rotations, one can create achromatic sequences. 

A third limitation is that the controlled polarisation rotations will differ slightly from $\pi$ rotations in a random way. Thus instead of implementing $\sigma_x$ one implements
$\exp[i (\pi \sigma_x + \vec r' \cdot \vec \sigma)/2]
= i\sigma_x (1 + i\vec r \cdot \vec \sigma + O(r^2))$ where $\vec r$ and $\vec r'$ are small random vector (which are simply related to each other). 
The main effect of these errors is that the stacks of 3 $\pi$ rotations mentioned above will no longer be perfectly achromatic. If we suppose that the random vectors $r_i$ associated with the successive controlled polarisation rotation are independent, then we 
 have to include an error term of the form
$O(\omega \beta \sqrt{N}r )$ where $N$ is the total number of controlled polarisation rotations and $r$ is the average length of the random deviations $r_i$.

As an illustration of the effects of imperfections let us consider the
case where the residual PMD is characterised by length scales $L_B \simeq L_{chge}\simeq 10m$. 
We correct for the PMD using the simple sequence of polarisation rotations eq. (\ref{seqsimple}), and in order to cancel the PMD of the polarisation rotations themselves, we replace each rotation by a triplet of rotations as described above. After one beat length the Jones matrix is of order
\begin{equation}
U(L_B) = 1 + O(\omega^0) + O(\omega B_1 l_{seq})
+ O( \beta \omega r \sqrt{L_B /l_{seq}} ) 
\ . \nonumber
\end{equation}
Assuming $r=1/30$, and inserting omitted numerical constants, we find that the PMD is reduced by a factor of $5$ for the optimal value of $l_{seq}=L_B/7$.


{\bf Conclusion.}
We have presented a method to reduce PMD in optical fibers based on the use of controlled polarisation rotations. The results presented here are only a first overview of the potentialities of the method, and we expect that significant improvements of the sequences presented can be achieved by 
using the same methods which are succesfully used to optimize pulse sequences in NMR.
The method presented here can be combined with the well established method  of spinning the fiber. The efficiency of the combined method will depend on the detailed properties of the residual birefringence in spun fibers, and on the precision with which the controlled polarisation rotations can be implemented.  In addition to the important application for long distance, high speed telecommunication in optical fibers, this method may find applications in other systems in which one wants to reduce unwanted birefringence. Finally, on the conceptual side, our work provides a simple system in which to test the ideas of bang-bang control.

We acknowledge financial support by the 
IAP
Programme - Belgium Science Policy - under grant V-18, and by the European Union projects RESQ and QAP.

\end{document}